\documentclass[twocolumn]{aastex631}

\usepackage{multirow}

\shorttitle{A Temperature Trend in Exoplanets Atmospheres}
\shortauthors{Estrela et al.}
\graphicspath{{./}{figures/}}

\begin{document}

\title{A Temperature Trend for Clouds and Hazes in Exoplanets Atmospheres}

\author{Raissa Estrela}
\affil{Jet Propulsion Laboratory, California Institute of Technology, 4800 Oak Grove Drive, Pasadena, California 91109, USA}

\author{Mark R. Swain}
\affil{Jet Propulsion Laboratory, California Institute of Technology, 4800 Oak Grove Drive, Pasadena, California 91109, USA}
 
 
\author{Gael M. Roudier}
\affil{Jet Propulsion Laboratory, California Institute of Technology, 4800 Oak Grove Drive, Pasadena, California 91109, USA}






\begin{abstract}
The transmission spectra of exoplanet atmospheres observed with the Hubble Space Telescope (HST) in the near-infrared range (1.1-1.65$\mu$m) frequently show evidence for some combination of clouds and hazes. Identification of systematic trends in exoplanet clouds and hazes is potentially important for understanding atmospheric composition and temperature structure. Here we report on the analysis of spectral modulation using a large, uniformly processed sample of HST/WFC3 transit spectra from 62 exoplanets. The spectral retrieval includes the capability to detect and represent atmospheres in which the composition departs from thermochemical equilibrium. By using this unique catalog and measuring the dampening of spectral modulations compared to strictly clear atmospheres, we identify two populations. One is completely cloud/haze free spanning a wide temperature range, while the other population, identified as ``Partial cloud/hazes'', follows a trend from mostly cloudy/hazy around 500~K to mostly clear at $\sim$1500~K. We also find that a partially transparent aerosol component is frequently present and that it is typically vertically distributed throughout the atmospheric column. Our findings also suggest that while clouds and hazes are common in exoplanet atmospheres, the majority of planets have some level of detectable spectral modulation. Additionally, the empirical trend that clouds and hazes are minimized at 1460.86K$^{+316}_{-405}$ revealed in our catalog has predictive utility for modelling the performance of large-scale transiting exoplanets survey, such as planned with the Ariel mission. This trend can also be used for making a probability-based forecast of spectral modulation for a given source in the context of future JWST observations. Future observations including the optical and/or a broader spectral coverage may be useful to further quantify the trend reported here.




\end{abstract}



\section{Introduction} \label{sec:intro}
NASA's Hubble Space Telescope (HST) has been used to observe exoplanet transits at wavelengths ranging from the ultraviolet (UV) to the visible to the near infrared (NIR) as long as 2.4~$\mu$m \citep{Charb02,Vidal03,Swain21,Deming13}. However, the majority of HST transiting-exoplanet spectroscopic observations have been conducted with the Wide Field Camera 3 (WFC3) instrument using the spatial-scan mode and the G141 grism covering wavelengths from 1.1~$\mu$m to 1.6~$\mu$m. Comparative planetology studies based on these observations show that the spectral modulation seen in the transmission spectra indicate atmospheres can vary from obscured to clear \citep{Sing16}. The nature of the obscuring aerosols, which we will term ``clouds and hazes'' is an area of active investigation, and possible candidates range from various condensed species to photochemically produced hazes. The cloudy and hazy atmospheres are associated with a weaker absorption feature in the NIR portion of the transmission spectrum of the planet or a featureless spectra \citep{Knutson14b,Knutson14a,Kreidberg14}. Although some of the low spectral amplitudes could also be associated with low water abundances, as has been pointed out by previous works \citep{Barstow17,Pinhas19, Wel19}. While in the optical spectra, clouds/hazes can cause Rayleigh scattering--like opacity seen in several planets \citep{Pont08, Pont13, Mc14, Sing16, Estrela21}. These clouds are generally attributed to atmospheric condensates of salt, silicate, and metal vapors \citep{Morley12,Zhang20,Gao20}, or to photochemically produced hazes due to energy input or energetic particle bombardment \citep{Lavvas17,Lavvas19,Gao20,Lavvas21}. Clouds and hazes can have a critical role in planetary atmospheres. They can affect the amount of light reflected by the atmosphere of the planet, which determines the Bond albedo and consequently the equilibrium temperature of the planet \citep{Cahoy10}. They can also control the thermal structure, chemistry, and dynamics in the atmosphere \citep{Tomasko10, Muller14}. 



A wide range of potential cloud compositions may be plausible due to the significant possibility of condensate materials \citep{Zhang20} and because of exoplanet properties, the potential for migration, and differences in formation/evolutionary scenarios. Particularly, cloud composition depends on the temperature of the atmosphere. Various scenarios using temperature dependence to infer cloud composition have been modeled \citep{Burrows99,Morley13,Parmentier16}. Some molecules will condense at high temperatures (e.g., TiO/VO, Al, and Ca), while others prefer moderate temperatures (e.g., Fe, Mg, Si, Cr, and Mn) or lower temperatures (e.g., K and Na) \citep{Gao21}. Observations of exoplanet atmospheres illustrate this diversity.  Many species, such as TiO/VO and metals, have been suggested to be present in ultrahot Jupiters \citep{Ben-Yami20,Cabot20,Chang22}, while smaller and cooler planets (sub-Neptunes and super-Earths) also have indicated the presence of high-altitude aerosols in their atmospheres \citep{Ehren14,Wake17b,Kreidberg20}. Other planets have shown very steep optical slope in their spectra \citep{Alderson20,Estrela21,Chen21} due to scattering in the atmosphere, which could be explained by sulphide clouds (MnS, ZnS, Na$_{2}$S) \citep{Pinhas17}. High-altitude photochemical hazes could also lead to super-Rayleigh optical slope, depending on their mixing and formation rate \citep{Ohno20}. Advances in laboratory experiments are helping to better constrain cloud formations, and simulations for hot Jupiters' atmospheres and even smaller planets' atmospheres have shown photochemical haze production \citep{Horst18,Fleury19,Fleury20,Yu21}. 

In addition to understanding cloud formation, the overall level of ``cloudiness'' in exoplanet atmospheres is a topic of great interest for planning exoplanet observations because of the potential for strong clouds to dampen spectral features and thus compromise the study of atmospheric composition \citep{Kreidberg14}. A robust method of predicting the likelihood and impact of clouds and hazes in the detection of exoplanet atmospheres is of great value for planning observations. For example, the Ariel mission \citep{Tinetti18,Zellem19} will conduct a transit spectroscopy survey of hundreds of exoplanets, and the ability to estimate cloud properties could be an important factor in planning the mission observations. Also, the detectability of molecular features on smaller planets with the James Webb Space Telescope (JWST) could be significantly compromised due to the presence of clouds/hazes \citep{Fauchez19}. Several studies have used HST/WFC3-G141 transmission spectra to conduct statistical analysis using the amplitude of the water feature to identify planets affected by the presence of clouds/hazes. Based on the HST/WFC3-G141 catalog of \cite{Tsiaras18}, \cite{Iyer16, Fu17,Fisher18, Gao20} have found that most planets had their amplitude diminished by aerosols when compared to a clear atmosphere. \citep{Wake19} show that 30\% of the time the amplitude of the water spectral feature is find to be below 0.5Hs which deviates from a cloud-free scenario. An analysis by \cite{Stevenson16} suggest that there is a threshold for cloudiness level at a surface gravity of 2.8~dex. More recently, \cite{Dymont21} have searched for correlations between cloudiness metrics and other physical parameters such as temperature using sample sizes of 23 planets. However, as noted by \cite{Dymont21}, the lack of a uniform data reduction was a less-than-ideal aspect of the sample they studied. 

Given the importance of clouds and hazes for establishing planet properties and their impact on characterization using the transit spectroscopy method, additional investigation into the properties of exoplanet clouds and hazes is a priority. Rapid progress in exoplanet observations is leading to the ability to improve sample size with respect to previous works. Here, we report on an analysis of clouds and hazes using the largest uniformly processed exoplanet transmission spectra catalog available in \cite{Roudier21}.

\section{Methodology}\label{sec2}

Detection of clouds and hazes is typically done through measuring the degree of spectral modulation \citep{Sing16}, which can then be compared to estimates of the theoretical possible values \citep{Iyer16,Gao20} or plotted against various physical parameters to search for correlations \citep{Stevenson16,Fu17,Fisher18,Gao20,Dymont21}. Here, we perform this analysis using the catalog from \cite{Roudier21} that consists of 62 planets observed with the WFC3 instrument onboard the HST using the G141 infrared (IR) grism (1.1--1.6~$\mu$m) spatial scan observations. Compared with previous works, this exoplanet catalog has three important advantages for the study of clouds. First, this catalog is the largest uniformly processed catalog of HST transmission spectra that has been published. In this case, the data have all been processed though the same pipeline, and they have all been analyzed with the same spectral retrieval code. A uniformly processed catalog is particularly important when performing comparative planetology because differences in planetary spectra can be attributed to differences in astrophysics rather than analysis methods. Second, this catalog is based on the EXoplanet CALIbration and Bayesian Unified Retrieval (EXCALIBUR) pipeline \citep{Swain21,Roudier21,Estrela21,Huber22}, which processes HST/WFC3-G141 data at the full spectral resolution of the instrument. Using the full G141 spectral resolution avoids the potential reduction in spectral feature amplitude that can be caused by additional spectral averaging. Third, the CERBERUS spectral retrieval code, used for modeling the transmission spectra in this catalog, is capable of modeling atmospheres assuming both thermal equilibrium chemistry (TEC) and disequilibrium chemistry (DisEQ). The majority of previous works interpreted planet populations observed with the HST/WFC3 instrument as being consistent only with bulk TEC conditions, while atmospheres can undergo DisEQ due to photochemistry, which can be associated with a water-depleted spectrum. The DisEQ atmosphere assumed by \cite{Roudier21} does not fully exclude water but includes water as a subdominant feature and brings the possibility of other dominant signatures such as CH$_{4}$, C$_{2}$H$_{2}$, HCN, and CO$_{2}$. \cite{Roudier21} identifies a third category of planets that could not be determined, based on the model selection criteria, to be either DisEQ or TEC. We term this category ``Indeterminate'' and divide this group into those that have spectral modulation greater than 1H$_{s}$ and those that do not. 

\begin{figure}[!ht]
\centering
\includegraphics[width=0.5\textwidth]{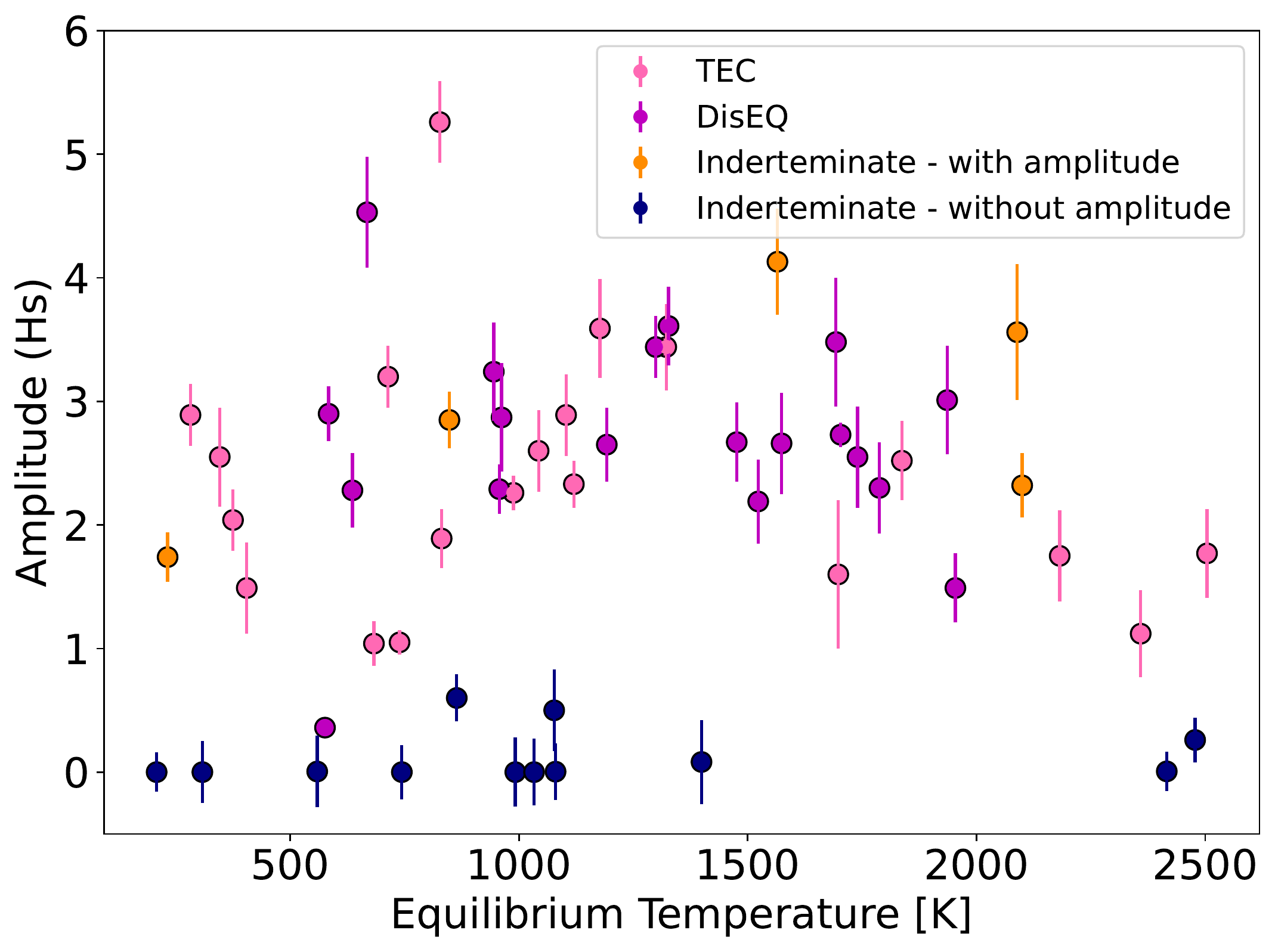}
\caption{Amplitude of the spectral modulation in TEC (light pink), DisEQ (magenta), and Indeterminate (orange) targets in terms of scale height.} 
\label{fig:spectralamp}
\end{figure}

To estimate the amplitude of the spectral modulation in the transmission spectra, we take the difference between the Maximum and Minimum of the CERBERUS preferred model (henceforth referred to as CMM) in the NIR range between 1.15--1.6~$\mu$m (see Fig.~\ref{fig:spectralamp}). This method is based on the approach of \cite{Iyer16} and has the advantage of naturally debiasing the impact of individual spectral channel values on estimates of spectral modulation. In the case of TEC models, the CMM will measure the strength of the water feature, while for the DisEQ planets it will measure the strength of the other possible dominant features (CH$_{4}$, C$_{2}$H$_{2}$, HCN, and CO$_{2}$). To facilitate the analysis, we convert the CMM to units of planetary scale height H$_{s}$:

\begin{equation}
    H_{s} = \frac{k_{B} T_{eq}}{\mu g},
\end{equation}

\noindent where $k_{B}$ is the Boltzmann constant, $\mu$ is the mean
molecular weight of an atmosphere in solar composition
($\mu$ = 2.3 amu), $g$ is the surface gravity, and $T_{eq}$ is the
calculated equilibrium temperature.

Next, we build an index of the spectral modulation that represents a fraction of the potential modulation a planet achieves. This index is obtained by dividing the difference between the CMM of the spectral modulation of the regular retrieved model (CMM$_{\rm CRB}$) by the CMM of the cloud/haze free model (CMM$_{\rm FREE}$) of each individual target:

\begin{equation}
    Spec_{\rm index} = \frac{\rm CMM_{\rm CRB}}{\rm CMM_{\rm FREE}}
\end{equation}

\begin{figure}[!ht]
\centering
\includegraphics[width=0.50\textwidth]{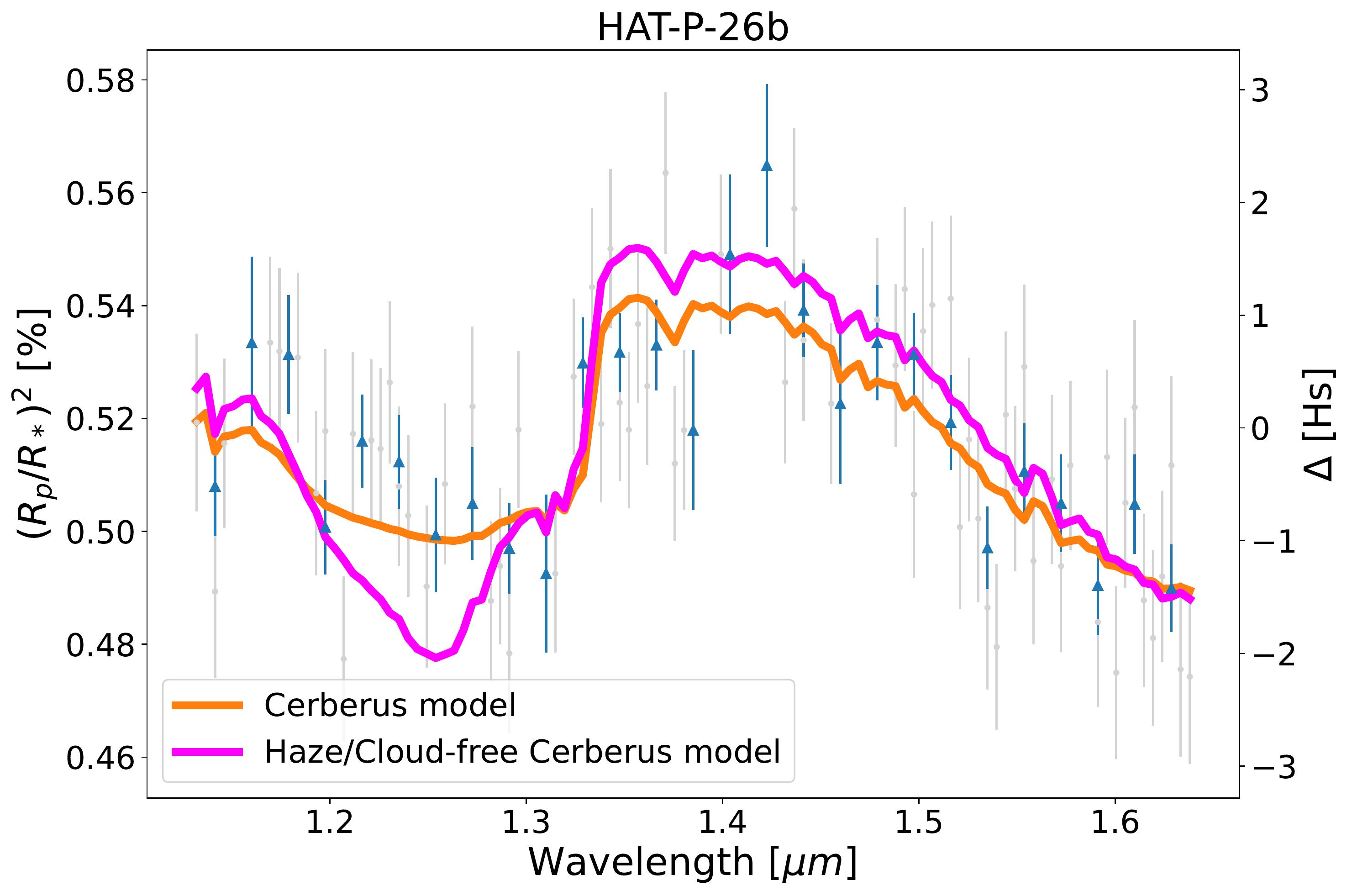}
\caption{Comparison between the retrieved Cerberus model for the target HAT-P-26b and the model without clouds and hazes used to obtain the spectral modulation index.}
\label{fig:spectralcomp}
\end{figure}

This index varies from 0 to 1, where 0 accounts for cloudy/hazy and 1 represents a clear atmosphere. The cloud/haze free model is obtained by setting the haze/clouds parameters to zero (see Fig. \ref{fig:spectralcomp} for comparison between the retrieved model and the cloud/haze-free model).



\subsection{Clouds and Hazes}

We also analyze the properties of the hazes and clouds obtained with the atmospheric retrieval CERBERUS. To derive information on the haze layer, CERBERUS uses a parametrization of Jupiter's haze profiles \citep{Zhang15} constructed by taking the median of the haze latitude number density. The following free parameters characterize the haze properties and are retrieved by CERBERUS:

\begin{itemize}
\item HScale: Multiplicative scale factor (log10) of the haze optical depth (common to all pressure layers). In CERBERUS, the optical depth per length element $\tau$/dl [m$^{-1}$] is given by:
\begin{equation}
   \rm \frac{\tau_{cerb}}{dl} = \rm HScale \times \rho_{cerb} \times \sigma_{cerb}
\end{equation}
where $\rm \rho_{cerb}$ and $\rm \sigma_{cerb}$ are the haze density profile and cross section in CERBERUS, respectively.
\item HLoc: The maximum density location of the haze cloud in log pressure (bars)
\item HThick: Multiplicative factor (log10) of the width of the density haze profile 
\end{itemize}

\begin{figure}[!ht]
\centering
\includegraphics[width=0.40\textwidth]{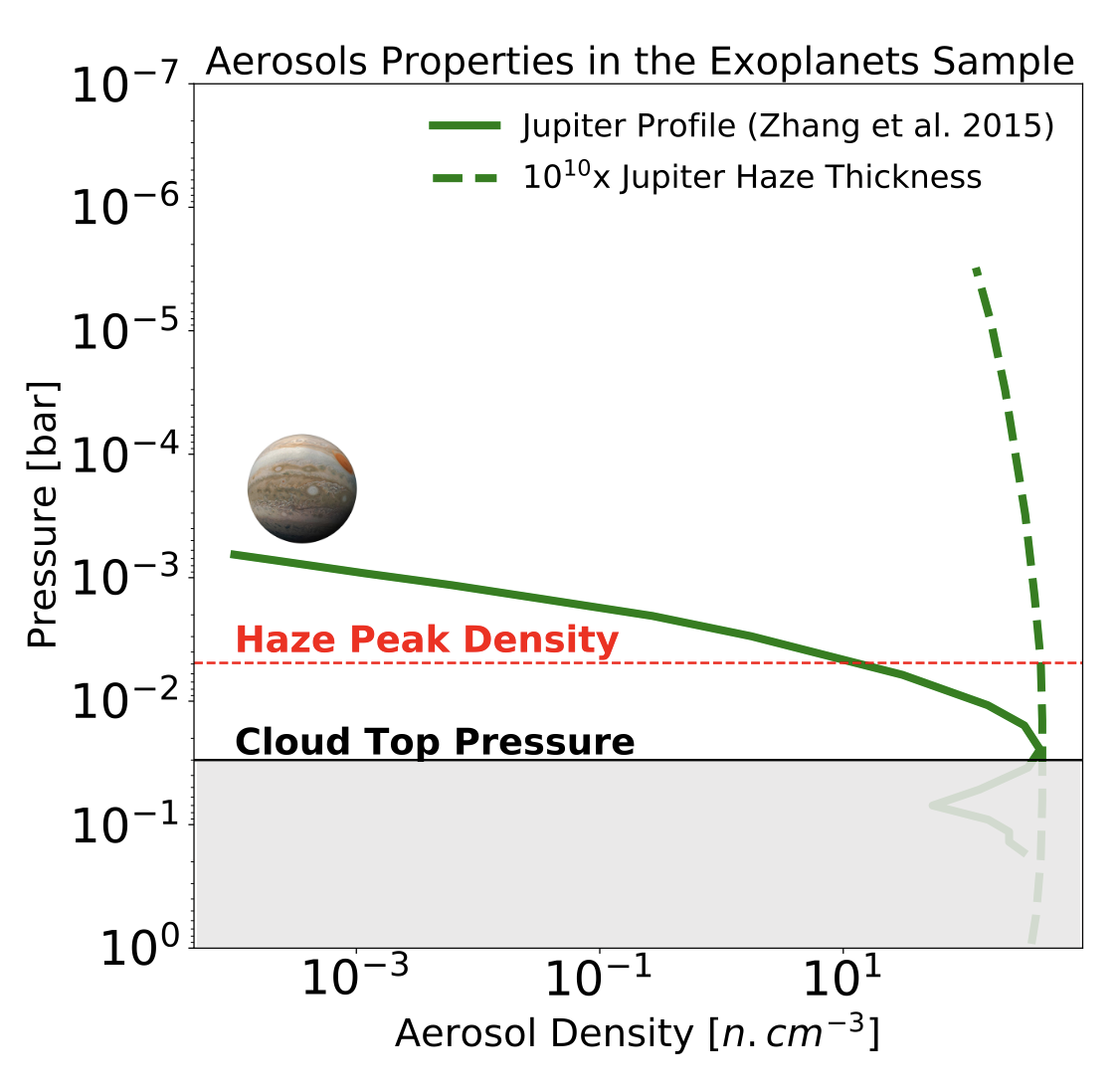}
\caption{Comparison between the Jupiter's haze profile and the scaled haze model obtained with the averaged aerosol properties found in the exoplanet sample. The thickness of Jupiter's haze layer is shown by a solid line and the median thickness found for the exoplanet catalog in this study is indicated by a dashed line. For the exoplanet model, we also illustrate the location of the haze peak density (in terms of pressure) in red and the cloud top pressure (CTP), indicated by a solid line. Any atmospheric content below the CTP (within the grey region) would be muted.}
\label{fig:aerosols}
\end{figure}

We illustrate in Fig.~\ref{fig:aerosols} how these parameters can vary for exoplanets. The aerosol density profile has a log-pressure maximum density location (HLoc) that is allowed to linearly translate vertically in the atmosphere. A multiplicative factor (HThick) is introduced to compress or stretch the vertical extent of the initial profile. Another multiplicative factor (HScale) is used to scale the haze optical depth for each pressure layer. In addition, the modeled atmosphere includes a bottom opaque cloud-top pressure (CTP) in a log unit of bars.

\section{Results}

The estimated amplitude of the spectral modulation of the TEC, DisEQ, and Indeterminate targets is shown in Fig.~\ref{fig:spectralamp}. The Indeterminate targets are classified into two groups: those with minimal to no spectral modulation and those with detected modulation $>=$1H$_{s}$. Targets in these groups are indecisive between TEC and DisEQ because the evidence of the model is not strong enough, and the model cannot decide between TEC or DisEQ to explain any modulation there. Fig.~\ref{fig:spectralamp} shows that our sample is very diverse in terms of equilibrium temperature. The distribution of the modulation varies from planets that have large modulation and are potentially cloud/haze free to those with minimal spectral modulation that could be in a completely aerosol-dominated scenarios or have high molecular weight atmospheres. See appendix for values of the spectral amplitude presented in Fig.~\ref{fig:spectralamp}.

\begin{figure*}[!htp]
\centering
\includegraphics[width=0.90\textwidth]{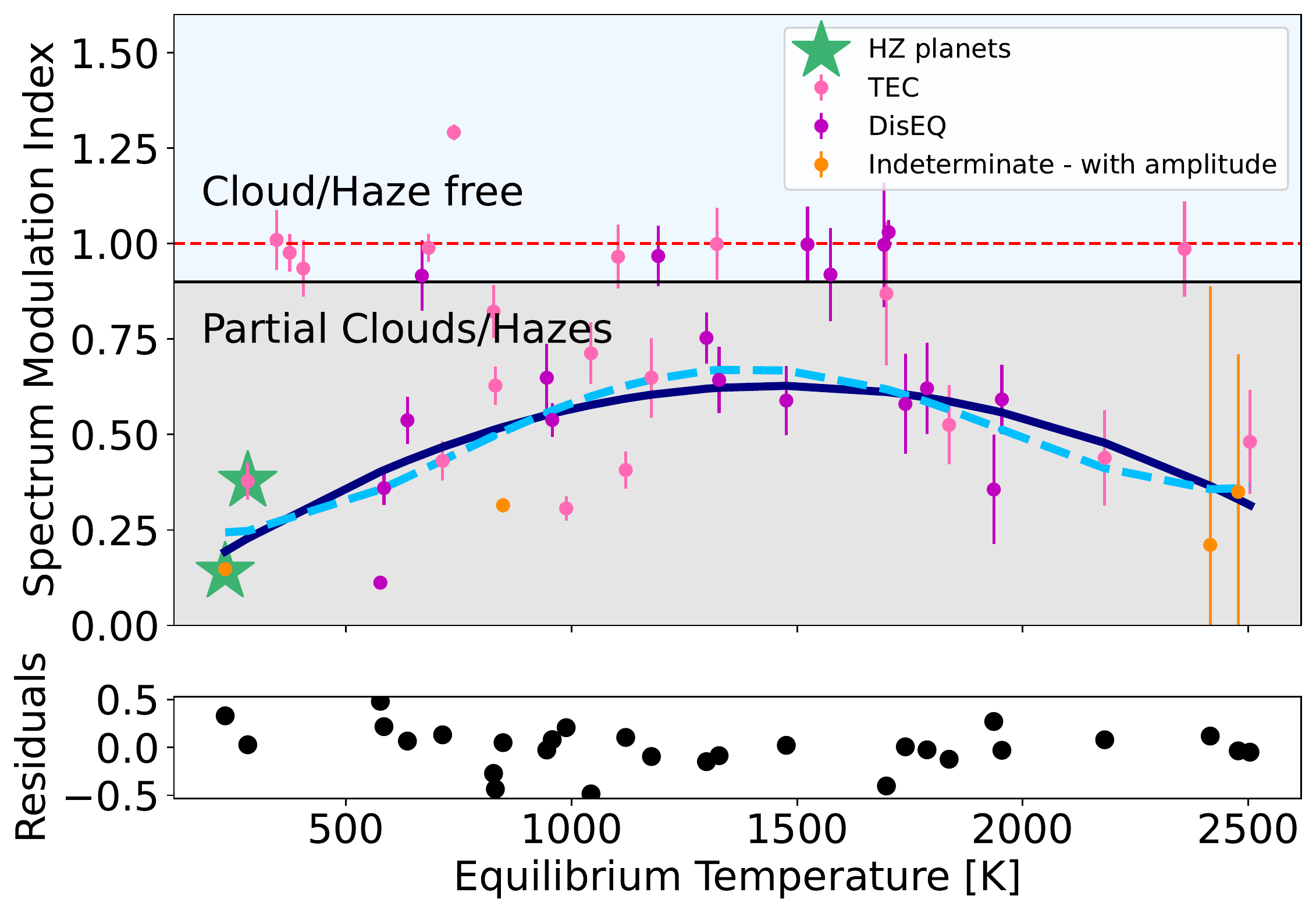}
\caption{[Top] Spectral modulation index of the TEC (light pink), DisEQ (magenta) and Indeterminate (orange) targets from the \cite{Roudier21} catalog gives the cloudiness level of each target. The solid blue line shows the best-fit for the ``Partial clouds/hazes" targets (within grey region), a second-order polynomial to the data that can be used as a forecast of spectral modulation to future missions. The dashed blue line shows a 4th order polynomial that highlights an increase in reduced hazy/cloudy atmospheres for the lower temperatures planets that are in the habitable zone. The blue shaded region indicates the zone of the targets with cloud/haze-free atmospheres. [Bottom] Residual of the spectral modulation index subtracted from the model.}
\label{fig:spectralindex}
\end{figure*}

The planets' range of cloudiness/haziness level is also well-illustrated in Fig.~\ref{fig:spectralindex} through the spectral modulation index (see description in Section \ref{sec2}). We identify two groups in Fig.~\ref{fig:spectralindex}. One group is cloud/haze-free (blue region) with spectral modulation indexes above 0.9 that is distributed across different temperatures. Within this group, there are a few ``Indeterminate" targets that have indexes greater than 1.3 due to steep slopes towards short wavelengths. We exclude these particular targets because their slopes could be associated to Rayleigh scattering by clouds/hazes or atmospheric escape. The second group has some level of cloudiness/haziness and we identified them as ``Partial clouds/hazes" region (in grey). If we exclude the completely cloud/haze-dominated (blue targets in Fig.~\ref{fig:spectralamp}) and the cloud/haze-free targets, the ``Partial clouds/hazes" targets show a rise in the spectral modulation index between 500--1500~K, indicating that we expect more clear atmospheres for planets with temperatures between $\sim$1200--1500~K. This trend is consistent with laboratory studies \citep{Yu21} that show that in the temperature range of 500--800~K, hazes can be removed more efficiently by direct deposition to the surface or deep atmosphere. Also, the trend seen from 500--1500~K is also in agreement with the cloudy-to-clear trend from 500-1000K seen in previous observational studies \cite{Crossfield17,Lilly20,Gui21} for warm Neptunes. Planets with higher temperatures ($>$ 1700K) seem to indicate more cloudy/hazy atmospheres although we lack sufficient data to make firm conclusions. However, it is possible that heat transport is very inefficient above 1600~K, which would allow the nightside to cool and clouds to advect to the dayside \citep{Parmentier21}. The temperature trend in the spectral modulation is well fitted using a second-order polynomial. The fit gives the relationship:

\begin{equation}
    y = -2.9e-7^{+7.1e-8}_{-7.1e-8}x^{2} + 8.4e-4^{+2e-4}_{-2e-4}x + 1.4e-2^{+0.13}_{-0.13}.
\end{equation}

If we fit a 4th order polynomial to the data (dashed blue line in Fig.~\ref{fig:spectralindex}), we see a trend raising towards reduced cloud/hazy atmospheres at low temperatures planets ($<$ 270K) in the Habitable Zone (HZ), K2-18b and LHS 1140b, where liquid water can exist and condense. This trend is also explained by laboratory studies \citep{Yu21}, which has shown that high surface energy hazes can be formed in these planets and they are efficiently removed by cloud droplets formed through water condensation. However, any conclusions can't be made due the lack of targets at lower temperatures. In addition, the 2nd order polynomial is still the best fit for the ``Partial clouds/hazes" population (although the 4th order is not excluded), as shown by the Bayesian information criterion (BIC). The BIC for each fit is calculated by: 

\begin{equation}
    {\rm BIC} = \chi^{2} + k \ln (n)
\end{equation}

where $k$ is the number of free parameters, represented by the polynomial order plus 1, and n is the number of data points. The best model is the one with the lowest BIC, which is the quadratic model, as shown in Table \ref{tab:bic}. We also estimated the delta BIC with respect to the best model, and found a value of 0.6 for the 4-th model which provides weak evidence that the quadratic model is better. On the other hand, the linear model has a delta BIC greater than 10, which represents strong evidence that the quadratic model is better.

\begin{table*}[!ht]
\centering
\begin{tabular}{ c c c c}
 Model & BIC & $\Delta$BIC & $\Delta$BIC rules of thumb\\ 
 \hline
 Linear & 190 & 70 & evidence against the linear model is strong\\  
 2nd order & 120 & 0 & best model \\
 3rd order & 123 & 3 &  evidence against the 3rd order polynomial is moderate\\
 4th order & 121 & 1 & weak evidence that best model is better\\
\end{tabular}
 \caption{Bayesian Inference Criteria (BIC) for each model fitted to the `Partial clouds/hazes" population.}
 \label{tab:bic}
\end{table*}



Considering that future missions would largely benefit from a forecast of spectral modulation for planning observations, we note that the second order polynomial trend can be used to make a spectral modulation forecast. Including the forecast for the temperature dependence for spectral modulation has the potential to increase the fidelity of planning observations with JWST and Ariel, and based on the results of this study, we would anticipate that plenty of targets would likely show modulation in their spectra.




\subsection{Aerosol Properties}

Another key result that comes from the analysis of the \cite{Roudier21} exoplanet catalog of TEC and DisEQ targets relies on the aerosol properties. To infer the aerosol properties, we take the best solution obtained with the CERBERUS atmospheric retrieval code by parametrizing the Jupiter aerosol layer (see Section~\ref{sec2}). We find that this sample of exoplanets has a much more extended haze layer than Jupiter, and it is widely distributed throughout the atmosphere. The CTP is very similar in both TEC and DisEQ targets, showing pressures of $\sim$0.05~bars, which means that any atmospheric column below that pressure is blocked by the cloud layer and not sampled by transmission spectroscopy. These results are summarized in Table~\ref{tab:aerosol} and illustrated in Fig.~\ref{fig:aerosols}, where the thickness of the haze layer, the location of the peak density, and the multiplicative factor of the haze opacity (with respect to Jupiter's opacity) are shown.

We caution that our haze model parameters are a parametric scaling of Jupiter's haze properties and thus do not capture the detailed microphysics of haze formation in these much hotter exoplanets. Our modeling suggests that haze in exoplanet atmospheres is typically distributed widely throughout the atmospheric column, similar to what was found for HD 209458b by \cite{Lavvas17} based on a microphysics model for haze formation. Similarly, our modeling suggests that the presence of a bottom cloud is detectable and influences the interpretation of WFC3/G141 observations. However, detailed physical conclusions about the vertical distribution of the haze and bottom cloud properties will need to be addressed with models that appropriately represent the chemistry and physics of these atmospheres---and we identify this as an area for further investigation.

\begin{table}[!ht]
\centering
\begin{tabular}{llll}
\multicolumn{4}{c}{\begin{tabular}[c]{@{}c@{}}Aerosol and Cloud Properties of the TEC and DisEQ\\ Exoplanets in the Sample\end{tabular}} \\
\hline
                      & TEC                                 & DisEQ                               & Indeterminate                        \\
HScale                & 0.008                               & 0.0006                              & 0.70                                 \\
HLoc                  & 0.005 bars                          & 0.006 bars                          & 0.003 bars                           \\
HThick                & 5.8 $\times$ 10$^{9}$               & 5.5 $\times$ 10$^{9}$               & 4.7 $\times$ 10$^{10}$               \\
CTP                   & 0.03 bars                           & 0.06 bars                           & 0.002 bars                               
\end{tabular}
\caption{Median values of the aerosol properties (HScale, HLoc, HThick) and the cloud top pressure (CTP) of the TEC and DisEQ population obtained with the atmospheric retrieval CERBERUS. }
\label{tab:aerosol}
\end{table}

\section{Discussion}\label{sec:dis}

\subsection{Cloudy/Hazy-to-Clear Atmospheres}
One of the key results from our analysis is that planets can take on a wide range of values from clear to cloudy atmospheres, but they are frequently somewhere in between, implying that the majority of the planets has some content of clouds and/or hazes. This result is consistent with previous work by \cite{Sing16} that shows that planets can have cloudy and/or hazy or clear atmospheres. Although some of the observed low spectral amplitudes \cite{Sing16} have been shown to be caused by low water abundances rather than by clouds/hazes exclusively \citep{Barstow17, Pinhas19, Wel19}. Our work also supports the claim that clouds/haze can indeed mute spectral features, as previously discussed by \cite{Iyer16}. However, in the majority of cases, the atmospheres still have detectable spectral features that are sufficient for detailed study, such as whether the atmosphere is TEC or DisEQ.

\subsection{Extended Aerosol Layer in Exoplanets}
We find that the planets in our catalog have a population of aerosol particles that are typically widely vertical through the exoplanet atmosphere ($\sim$4 orders of magnitude in pressure). This wide distribution of particles raises questions about whether they could be condensates, as the T-P profile could vary considerably over this pressure range. Another possible explanation could be photochemical haze particles, as they have been shown to be distributed throughout the atmospheric column of the planet HD~189733b extending from 10$^{2}$~bar to millibars \citep{Lavvas17}. However, the composition of the clouds is not explored in detail in this work because condensation of materials is strongly dependent on the temperature, and our atmospheric temperature is obtained with a 1D atmospheric retrieval, which in some cases can lead to biased values of the temperature of the atmosphere \citep{Wel22}. Therefore, further investigation is needed to determine the nature of the aerosol particles. 

\subsection{Forecasting Spectral Modulation}
Using the spectral amplitudes seen in our catalog, we estimate a spectrum modulation index that varies from 0 to 1, where 0 would be a cloudy atmosphere with no spectral amplitude and 1 would be a cloud-free atmosphere. We found a quadratic polynomial trend in the spectrum modulation index of the targets within the partial clouds/hazes region that can be used for forecasting cloudiness/haziness for a given source that could be potentially observed by current or future missions. By taking the derivative of this equation, we get that the minimum clouds/haze occurs at 1460.86$^{+316}_{-405}$~K. The empirically determined trend is useful for forecasting the level of spectral modulation that will be detected in large-surveys, such as the one planned with the Ariel mission. Our findings suggest that additional time in the form of repeated visits should be considered for observations of lower temperature planets to increase the signal to noise of the transit spectrum to compensate for the reduction in spectral modulation caused by clouds and hazes. A larger sample such as the one anticipated by Ariel will certainly be useful for improving this type of empirical constraint on cloud/haze impacts on spectral modulation.

\begin{table*}[!ht] 
\centering 
\begin{tabular}{ccc} 
\hline
Step & Dependence Chain & Reference Used \\ 
\hline 
Make Spectra & 4 pixels spectra averaging & \cite{Tsiaras18}  \\ 
\multirow{2}{*}{Model Spectra} & TEC only model             & \multirow{2}{*}{\cite{Fu17} }    \\ 
             &  H$_{2}$O as the dominant opacity source            & \\
Estimate Modulation  & Max/Min of best fit model at 1.3-1.65$\mu$m & \cite{Fu17}  \\
\hline 
Step          & Dependence Chain in this work & Reference Used  \\
\hline 
Make Spectra  & No Averaging & \cite{Roudier21} \\
\multirow{2}{*}{Model Spectra} & TEC and DisEQ models & \multirow{2}{*}{ \cite{Roudier21}}  \\
             &  Many molecular species  & \\
Estimate Modulation  & Max/Min of probable model at 1.1 and 1.6$\mu$m  & \cite{Roudier21} \\
\hline
\end{tabular}
  \caption{Summary of the key differences between the methods used in previous works in the literature and those used in this work.} 
  \label{tab:tab2} 
\end{table*} 

\subsection{Differences Between EXCALIBUR and Previous Works}

In addition to conducting an analysis using a large, uniformly observed and processed catalog of transit spectra, there are two key aspects of our analysis. The first is that the transit spectra in this catalog are processed at the full spectral resolution of the HST/WFC3 instrument. The key advantage here is that it minimizes the potential for reducing spectral modulation by low-pass filtering. A previous sample analyzed by \cite{Fu17} did not use a full resolution of the spectrum as EXCALIBUR does. Their sample was based on the data reduction by \cite{Tsiaras18} that used a 4-pixel spectral averaging while our sample does not average spectral bins. Similarly, \cite{Edwards22} applied the methods of \cite{Tsiaras18} and \cite{Fu17} to a larger sample and find qualitatively consistent results to \cite{Fu17}. Spectral averaging has the clear potential to decrease spectral modulation. The second key aspect is that the spectral retrieval analysis for the EXCALIBUR catalog can model both TEC and DisEQ atmospheric composition scenarios, including atmospheres that show spectral modulation due to water vapor. Thus, unlike some previous work \citep{Fu17,Fisher18,Gao20,Dymont21,Edwards22}, our approach can measure and model spectral modulation independent of whether a water signature is present in the spectrum. Assuming only water to be the possible opacity in the spectrum has the potential to force a water-model fit to a water-depleted atmosphere, which could result in lower spectral modulation. This constitutes a substantial difference from the approach taken previously and is a point that needs to be borne in mind when comparing our results to previous work. 

Another difference between our catalog and previous work relies on the estimation of the spectral modulation. Here, we take the difference between the maximum and minimum of the most probable CERBERUS model in that interval, while \cite{Fu17} takes the difference between the maximum and minimum of the best-fit water model. All these factors together, summarized in Table \ref{tab:tab2}, could explain the difference in trend seen in the amplitude of the spectral modulation between our work and that of the previous sample by \cite{Fu17}, with the average amplitude found by EXCALIBUR being higher.

However, similar to \cite{Dymont21}, planets in our catalog that are below 1000~K can vary widely in terms of cloudiness around the overall trend. Also, our work seems to quantitatively agree with the quadratic polynomial trend seen in \cite{Yu21} that shows an increase in clear atmospheres from 500~K to 800~K. The inclusion of visible wavelengths would have the potential to reveal a better picture of the dominance of haze and clouds in the atmospheres, as has been explored by previous works \citep{Heng16, Pinhas19}. Therefore, future catalogs may benefit from including the spectral modulation from observations taken by the Space Telescope Imaging Spectrograph (STIS) on the HST. However, the merge of HST/STIS and HST/WFC3 can be challenging since there are multiple factors that could compromise the interpretation: (1) the average planet/star ratios measured with HST/STIS can be different from than that obtained with HST/WFC3, and this difference could be associated with intrinsic properties of the atmosphere, or due to different instrument systematics or stellar activity \citep{Estrela21}; (2) effects from unocculted star spots that could produce slopes in the optical transmission spectrum \citep{Rack17}.

\section{Conclusions}

This work performs comparative planetology of a large, uniformly observed and processed catalog of transit spectra that were obtained at the full spectral resolution of the HST/WFC3 instrument. Our method is consistent for a large range of planet properties, and the sample is larger than those used in previous works. Another differentiating aspect of this work when compared with previous studies is that the spectral catalog used here is modeled with both TEC and DisEQ atmospheric composition scenarios. This implies that the spectral modulation in exoplanet spectra is modeled assuming not only H$_{2}$O but also other dominant spectral features due to disequilibrium chemistry. The amplitude of the spectral modulation in our catalog reveals that exoplanet atmospheres vary from cloudy to clear, with most of the planets typically lying in between. This suggests that transiting exoplanet atmospheres contain detectable spectral features that are partially muted. From the analysis of the haze properties, we find that haze layer is partially transparent and widely extended throughout the atmospheric column. 

In addition, we estimated a spectral modulation index that quantifies the cloudiness/haziness in each target. We observe that there are two groups of planets, one is completely cloud/haze free spanning a wide temperature range. The other group, denominated ``Partial cloud/haze", follows a trend from cloudy/hazy to clear within 500--1500~K in the spectral modulation, with the least cloudy atmospheres lying at 1460.86K$^{+316}_{-405}$. This trend is consistent with previous observational samples and laboratory studies from \cite{Yu21} that indicate efficient haze removal in that temperature range. We find that a quadratic polynomial is the best fit for the trend seen in the spectral modulation indexes. However, a 4th order polynomial trend shows a rise of reduced hazy/cloudy atmospheres for planets with temperatures lower than 270K, including those in the HZ. At these temperatures, water can condense and hazes can be removed by wet deposition \citep{Yu21}. The empirical quadratic polynomial trend can be used for predicting cloudiness/haziness for modeling large-scale surveys. This is particularly beneficial for forecasting potential targets that could have their atmospheres observed by JWST or future missions devoted to the the observation of exoplanet atmospheres, such as Ariel. A potentially more robust improvement in the estimation of the spectral modulation index could be achieved by including the measurements on the optical wavelengths from HST/STIS in future catalogs. HST/STIS can give information on scattering due to haze and clouds, and would provide a more definitive picture of the range of cloudiness in exoplanet atmospheres.

\section*{Acknowledgments}

We would like to thank the anonymous reviewer for their constructive feedback that helped to improve this manuscript. We also thank Robert West for useful comments that provided a valuable addition to improve this manuscript. This research has made use of the NASA Exoplanet Archive, which is operated by the California Institute of Technology under contract with the National Aeronautics and Space Administration under the Exoplanet Exploration Program. Raissa Estrela, Mark Swain, and Gael Roudier acknowledge support for a portion of this effort from NASA ADAP award 907524. This work has been supported in part by the Jet Propulsion Laboratory, California Institute of Technology Exoplanet Science Initiative. This research was carried out at the Jet Propulsion Laboratory, California Institute of Technology, under a contract with the National Aeronautics and Space Administration (80NM0018D0004). ©2022 California Institute of Technology. Government sponsorship acknowledged.

\bibliography{sample631}{}
\bibliographystyle{aasjournal}



\appendix
\label{appendixA}

\section{Supplementary Material}

\begin{table}[!ht]
\centering
\begin{tabular}{ccc}
\multicolumn{3}{c}{Spectral Modulation Amplitude -- \textbf{TEC}} \\
\hline
Planet         & T$_{eq}$ [K]    & Amplitude {[}Hs{]}    \\
\hline
GJ 3470b       & 683            & 1.04 $\pm$ 0.18       \\
HAT-P-11b      & 831            & 1.89 $\pm$ 0.24       \\
HAT-P-26b      & 1043           & 2.60 $\pm$ 0.33       \\
HAT-P-32b      & 1837           & 2.52 $\pm$ 0.32       \\
K2-18b         & 282            & 2.89 $\pm$ 0.25       \\
K2-3c          & 375            & 2.04 $\pm$ 0,25       \\
WASP-107b      & 739            & 1.05 $\pm$ 0.10       \\
WASP-12b       & 2504           & 1.77 $\pm$ 0.36       \\
WASP-17b       & 1698           & 1.60 $\pm$ 0.60       \\
WASP-39b       & 1120           & 2.33 $\pm$ 0.19       \\
WASP-6b        & 1103           & 2.89 $\pm$ 0.33       \\
WASP-69b       & 988            & 2.26 $\pm$ 0.14       \\
WASP-76b       & 2182           & 1.75 $\pm$ 0.37       \\
WASP-80b       & 827            & 5.26 $\pm$ 0.33       \\
XO-1b          & 1177           & 3.59 $\pm$ 0.40       \\
HAT-P-1b       & 1322           & 3.44 $\pm$ 0.35       \\
HD 97658b      & 714            & 3.20 $\pm$ 0.25       \\
TRAPPIST-1b    & 405            & 1.49 $\pm$ 0.37       \\
WASP-121b      & 2359           & 1.12 $\pm$ 0.35      
\end{tabular}
\label{tab:ap1}
\setcounter{table}{0}
\renewcommand\thetable{\Alph{section}.\arabic{table}}
\caption{Spectral Modulation Amplitude obtained with the \cite{Roudier21} catalog of the TEC planets.}
\end{table}

\begin{table}[!ht]
\centering
\begin{tabular}{cccll}
\multicolumn{3}{c}{Spectral Modulation Amplitde -- \textbf{DisEQ}} &  &  \\
\hline
Planet            & T$_{eq}$ [K]      & Amplitude [Hs]            &  &  \\
\hline
GJ 1132b          & 584           & 2.90 $\pm$ 0.22       &  &  \\
GJ 1214b          & 576           & 0.36 $\pm$ 0.04      &  &  \\
GJ 436b           & 636           & 2.28 $\pm$ 0.3       &  &  \\
HAT-P-12b         & 957           & 2.29 $\pm$ 0.2       &  &  \\
HAT-P-17b         & 945           & 3.24 $\pm$ 0.4       &  &  \\
HAT-P-41b         & 1936          & 3.01 $\pm$ 0.44      &  &  \\
HD 189733b        & 1192          & 2.65 $\pm$ 0.3       &  &  \\
HD 209458b        & 1476          & 2.67 $\pm$ 0.32      &  &  \\
KELT-11b          & 1703          & 2.73 $\pm$ 0.1       &  &  \\
WASP-29b          & 962           & 2.87 $\pm$ 0.44      &  &  \\
WASP-31b          & 1574          & 2.66 $\pm$ 0.41      &  &  \\
WASP-52b          & 1299          & 3.44 $\pm$ 0.25      &  &  \\
WASP-63b          & 1523          & 2.19 $\pm$ 0.34      &  &  \\
WASP-74b          & 1788          & 2.30 $\pm$ 0.37       &  &  \\
WASP-79b          & 1740          & 2.55 $\pm$ 0.41      &  &  \\
55Cnc e            & 1954          & 1.49 $\pm$ 0.28      &  &  \\
GJ 9827b          & 668           & 4.53 $\pm$ 0.45      &  &  \\
XO-2b             & 1327          & 3.61 $\pm$ 0.32      &  &  \\
HD 149026b        & 1693          & 3.48 $\pm$ 0.52      &  &  \\
TRAPPIST-1c       & 346           & 2.55 $\pm$ 0.4       &  & 
\end{tabular}
\renewcommand\thetable{\Alph{section}.\arabic{table}}
\label{tab:ap2}
\caption{Spectral Modulation Amplitude obtained with the \cite{Roudier21} catalog of the planets under DisEQ.}
\end{table}

\begin{table}[!ht]
\centering
\begin{tabular}{ccc}
\multicolumn{3}{c}{Spectral Modulation Amplitude -- \textbf{Indeterminate}}                                 \\
\hline
Planet                       & Teq {[}K{]}              & Amplitude {[}Hs{]}            \\
\hline
GJ 3053b                     & 232                      & 1.74 $\pm$ 0.2                   \\
HAT-P-18b                    & 848                      & 2.85 $\pm$ 0.23                  \\
HAT-P-38b                    & 1077                     & 0.50 $\pm$ 0.33                   \\
HATS-7b                      & 1080                     & 0.00 $\pm$ 0.23                \\
K2-24b                       & 744                      & 0.00 $\pm$ 0.22                     \\
K2-3d                        & 308                      & 0.00 $\pm$ 0.25                     \\
K2-33b                       & 864                      & 0.60 $\pm$ 0.19                   \\
K2-96c                       & 559                      & 0.00 $\pm$ 0.29                 \\
KELT-1b                      & 2416                     & 0.00 $\pm$ 0.16                 \\
KELT-7b                      & 2089                     & 3.56 $\pm$ 0.55                  \\
Kepler-16b                   & 208                      & 0.00 $\pm$ 0.16                     \\
WASP-101b                    & 1565                     & 4.13 $\pm$ 0.43                  \\
WASP-18b                     & 2478                     & 0.26 $\pm$ 0.18                  \\
WASP-19b                     & 2100                     & 2.32 $\pm$ 0.26                  \\
WASP-62b                     & 1399                     & 0.08 $\pm$ 0.34                 \\
WASP-67b                     & 1033                     & 0.00 $\pm$ 0.27                  \\
\end{tabular}
\renewcommand\thetable{\Alph{section}.\arabic{table}}
\label{tab:ap3}
\caption{Spectral Modulation Amplitude obtained with the \cite{Roudier21} catalog of the planets that could not be determined, based on model selection criteria of the the atmospheric retrieval, to be either DisEQ or TEC.}
\end{table}

\end{document}